\documentclass[1p]{elsarticle}

\usepackage{bm}
\usepackage{notes2bib}
\usepackage{graphicx}
\usepackage{mathtools}
\usepackage{amsmath,amsbsy,amssymb,amsfonts}  

\newcommand*{\half}{\frac{1}{2}}

\newcommand*{\ket}[1]{\lvert #1 \rangle}

\newcommand*{\Evec}{\vec{\mathbf{E}}}

\biboptions{sort&compress}

\begin{document}
\begin{frontmatter}

\title{Optical bistability in a $\Lambda$-type atomic system including near 
dipole-dipole interaction}

\author{Juan D. Serna\corref{cor1}}
\ead{juan.serna@scranton.edu}
\address{Department of Physics and Electrical Engineering, University of
         Scranton, Scranton, Pennsylvania 18510, USA}
 
\author{Amitabh Joshi\corref{cor2}}
\ead{mcbamji@gmail.com}
\address{Department of Physics and Optical Engineering,
         Rose-Hulman Institute of Technology, Terre Haute, Indiana 47803, USA}
 
\cortext[cor1]{Corresponding author}
\cortext[cor2]{Principal corresponding author}

\begin{abstract}
The advantage of optical bistability (OB) using three-level electromagnetically 
induced transparency (EIT) atomic system  over the two-level system is its 
controllability, as absorption, dispersion, and optical nonlinearity in one of 
the atomic transitions can be  modified considerably by the field interacting 
with nearby atomic transitions. This is due to induced atomic coherences 
generated in such EIT system. The inclusion of near dipole-dipole (NDD) 
interaction among atoms further modifies absorption, dispersion, and optical 
nonlinearity of three-level EIT system and the OB can also be controlled by this 
interaction, producing OB to multistability.
\end{abstract}

\end{frontmatter}

\section{\label{sec:intro}Introduction}

Atomic optical bistability (AOB) is a phenomenon that exploits both the 
cooperative nature of the interaction between a group of atoms with the field 
and its strong nonlinearity. The essential element in AOB is the nonlinear 
relation between the applied electromagnetic field and the electromagnetic field 
that is radiated by the atomic dipoles, when the actions of all the other 
dipoles are taken into account. Cooperative radiation therefore plays a key role 
in bistability~\cite{Mandel}.

The optical bistability (OB) in an ensemble of two-level atomic systems 
contained inside an optical cavity has been extensively studied in the decade of 
eighties and early years of nineties. The studies were focused mainly on to 
utilize the phenomenon of OB in optical memories, optical transistor-like 
systems and all-optical switches~\cite{Lugiato84,Gibbs}. The observed phenomenon 
of OB is classified as absorptive OB if the atomic transition is saturated. On 
the other hand the intensity-dependent refractive index of the media is the 
mechanism for the generating dispersive 
OB~\cite{Lugiato84,Gibbs,Gibbs76,Venkatesan77,Agrawal79,Boyd}.

OB is a phenomenon that exploits both the cooperation among atoms and the strong 
nonlinearity produced by the interacting field. The two-level OB systems show 
first-order phase transition like behavior because in Focker–Plank equation the 
diffusion coefficient is intensity dependent~\cite{Lugiato80,Lugiato81}. Several 
interesting theoretical and experimental works reported for the two-level OB 
systems in past are well documented in the literature. However, experiments with 
two-level OB systems have limitations due to the lack of control in generating 
desirable OB curves with the help of only one field. The electromagnetically 
induced transparency (EIT) phenomena and related quantities, e.g., dispersion, 
nonlinearity etc. in multi-level systems have given a significant impact on the 
control of OB and optical multistability in such 
systems~\cite{Arimondo96,Harris97,Marangos98,Boller91,Field91,Li95,Gea95:51, 
Xiao95} . In multi-level systems, the two or more than two interacting fields 
are generating the induced atomic coherence, which modifies the linear 
absorption, dispersion and enhances the third-order Kerr nonlinear index of 
refraction.

Some studies on the behavior of nonlinearity with intensities and frequency 
detunings of both probe and coupling beams were reported~\cite{Wang01} in the 
$\Lambda$-type atomic system to demonstrate the controllability of OB during 
experiments in such system. Controlling of OB hysteresis cycle is useful in 
all-optical switches, optical memories, optical transistors, and logic circuits. 
 This will avoid optical–electronic–optical conversion of information in the 
form of signals. The atomic coherence in EIT systems enhances Kerr nonlinearity, 
and this brings down the switching thresholds of such devices. Thus the control 
becomes easier and more efficient, at very low intensity levels of light.

The cavity output field of atom-field composite system for some parametric 
conditions becomes unstable, giving rise to dynamic instability and chaos as the 
optical pumping in the coupling transition competes with the saturation in the 
probe transition~\cite{Yang04,Yang05}. This is achieved by adjusting the 
coupling beam Intensity and/or frequency detuning. The phenomenon of stochastic 
resonance~\cite{Joshi06:74} in optical bistability is also reported in 
three-level systems. The noise induced switching 
phenomenon~\cite{Joshi08,Joshi06:49} in three-level bistable systems is also 
observed, not available with the two-level systems conveniently.

Propagation of electromagnetic field can generate near dipole-dipole (NDD) 
interaction in a collection of atoms. The NDD effects can introduce the 
inversion-dependent-chirping of the atomic resonance frequency in a two-level 
atomic sample. The microscopic field-atom coupling can be modified due to NDD 
effect. The microscopic field results from the composition of macroscopic field 
with the induced polarization~\cite{Bowden93}. The NDD interaction is realized 
in a dense medium from entities confined in a very small volume of the order of 
a cubic wavelength. The modified Maxwell-Bloch equation in which NDD effects are 
included gives some extraordinary results such as invariant pulse propagation 
that departs from the hyperbolic secant pulse shape of self-induced transparency 
(SIT)~\cite{Bowden91} and self-phase modulation in SIT~\cite{Stroud88}. The 
interaction of external driving field with a atomic sample generates reaction 
field that works against driving field causing reduction of total field. The 
effect of strong external driving field in comparison to generated reaction 
field due to the dipole-dipole interaction, results in a first-order phase 
transition far away from the equilibrium due to the suppression of reaction 
field~\cite{Inguva90,Ben-Aryeh87}. The Bose Einstein condensates can be realized 
at high temperatures~\cite{Yang07} with the help dipole-dipole interaction. In 
systems used for quantum information processing the entanglement of 
qubits~\cite{Brennen00} can be achieved through dipole-dipole interaction. The 
dipole-dipole interaction is also important during the photosynthesis 
process~\cite{El-Ganainy13,Mohseni08}. The desired value of dipole–dipole or NDD 
interaction among entities can be achieved by change of the number density. The 
dipole–dipole interaction strength and its direction depends on environmental 
electromagnetic field modes, which can be obtained by directional stop-gaps or 
manipulating degeneration of the field modes~\cite{Kobayashi95,Agarwal98}. By 
using specific optical confinement structures, the desired dipole–dipole 
interaction can be realized~\cite{Hopmeier99,Levene03}.

Recently, the dynamical evolution of a three-level $\Lambda$-type atomic system 
including near-dipole–dipole interaction was investigated. The linear and third 
order non-linear susceptibilities for the probe transition was studied and 
details of absorption and dispersion in the probe transition and dynamic 
evolution of the system were obtained~\cite{Joshi19}. The aim of this work is to 
study the phenomenon of OB in a three-level $\Lambda$-type system including NDD 
interaction among atoms and thus demonstrate the controllability of OB via NDD 
interaction.

The remaining paper is organized as follows. In Section~\ref{sec:model} the 
model for this work for OB is described. Numerical results for OB are discussed 
in Section~\ref{sec:results} by the solution of the exact density matrix 
equation. In Section~\ref{sec:conclusion} some concluding remarks are presented.

\section{\label{sec:model}The Model}

\begin{figure}
  \centering
  \includegraphics[scale=0.9]{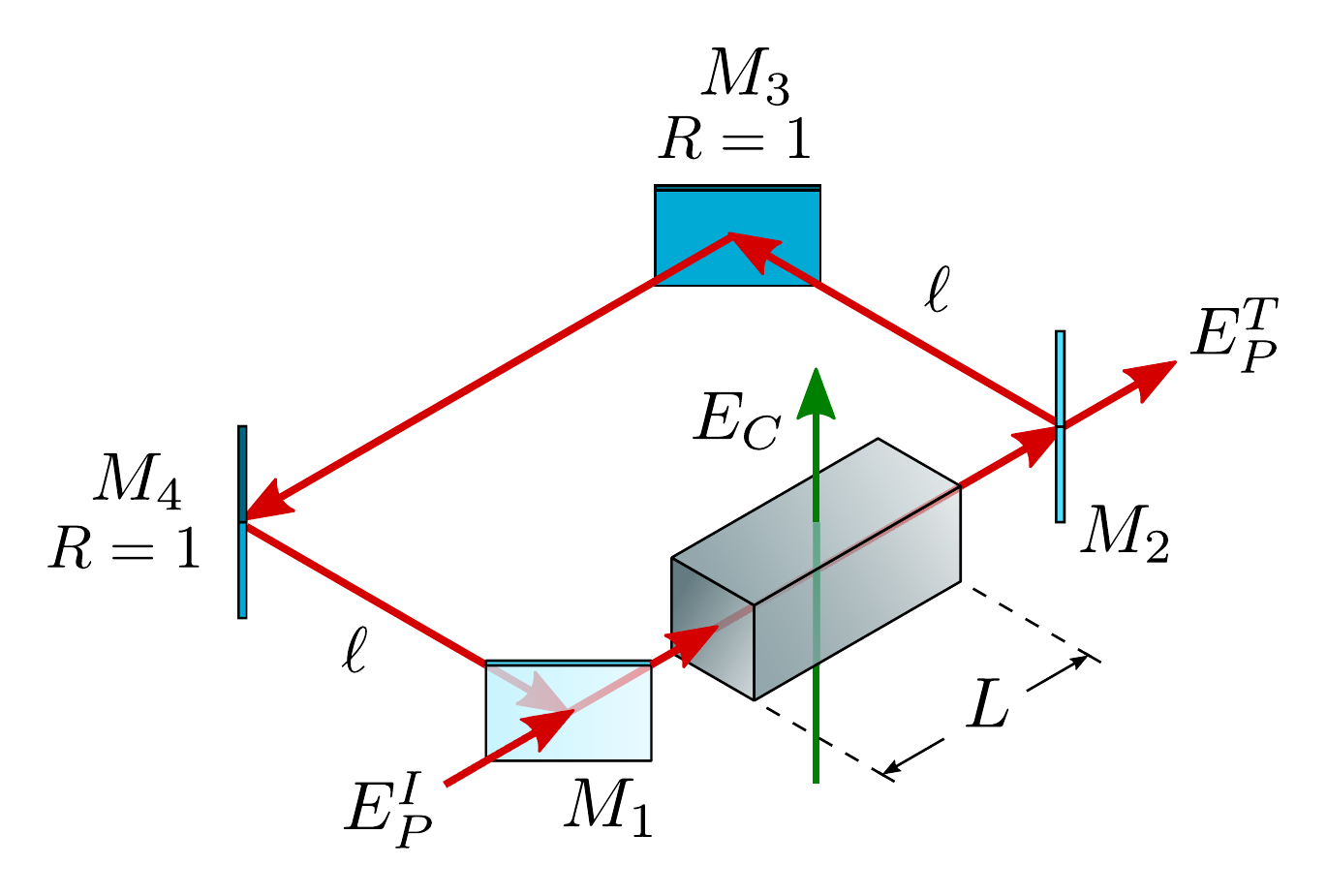}
  \caption{Schematic diagram of a unidirectional ring cavity having four mirrors
  ($M_1\text{--}M_4$) and an atomic vapor cell of length $L$. Mirrors $M_3$ and
  $M_4$ are perfectly reflecting mirrors ($R=1$). The incident and transmitted
  fields are represented by $E_P^I$ and $E_P^T$, respectively, while the 
  coupling field is represented by $E_C$.}
  \label{Fig:ring}
\end{figure}

\begin{figure}
  \centering
  \includegraphics[scale=1]{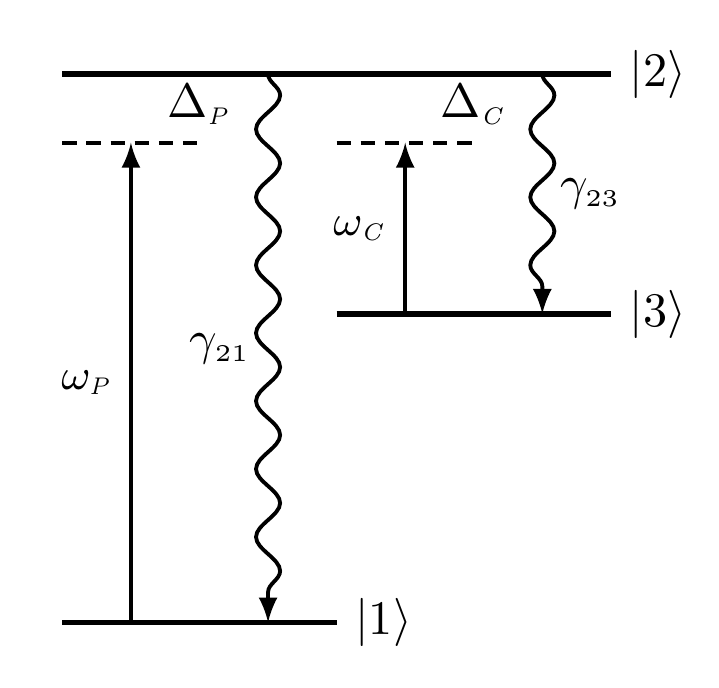}
  \caption{Schematic diagram of the three-level system in a
  $\Lambda$-configuration driven by probe and coupling lasers of frequencies
  $\omega_P$ and $\omega_C$, respectively.}
  \label{Fig:2}
\end{figure}

We consider a cavity illustrated in Fig.~\ref{Fig:ring}, in which there is a 
nonlinear medium of length $L$ in one arm, driven by a field $E_P^I$ that 
travels the medium in one direction only~\cite{Joshi06:49,Joshi10}. We assume 
that the input and output mirrors, $M_1$ and $M_2$, have identical reflectance 
$\sqrt{R}$ and transmittance $\sqrt{T}$ amplitudes (with $T=1-R$), whereas the 
other two mirrors are fully reflecting ($R=1$). We also assume that the mirror 
separation $\ell$ is adjusted so that the cavity is tuned to near resonance with 
the applied fields. The nonlinear medium consists of a three-level atomic system 
with energy levels $E_i$ ($i=1,2,3$) and in a $\Lambda$-type configuration ($E_2 
> E_3 > E_1$), interacting with probe and coupling lasers of frequencies 
$\omega_P$ and $\omega_C$, with amplitudes $E_P$ and $E_C$, respectively (see 
Fig.~\ref{Fig:2}). The probe laser acts on the dipole-allowed transition 
$\ket{1}\rightarrow\ket{2}$, with energy difference $\omega_{21}$ and frequency 
detuning $\Delta_P = \omega_P - \omega_{21}$. The coupling laser couples to the 
other dipole-allowed transition $\ket{2}\rightarrow\ket{3}$, with energy 
difference $\omega_{23}$ and frequency detuning $\Delta_C = \omega_C - 
\omega_{23}$.

In the semiclassical approximation, where the system interacts with the 
classical electromagnetic fields of the two lasers, the Liouville equations of 
the density-matrix elements in the dipole and rotating wave approximation, 
including NDD interaction as described in~\cite{Joshi19}, have the form
\begin{subequations}\label{Eq:DressedStates}
\begin{align}
  \dot{\rho}_{22} - \dot{\rho}_{11} =&  \notag
    - (\gamma_{23} + 2\gamma_{21})\rho_{22}
    + 2i\mu_{12}(\varepsilon_L^P)^*\rho_{21}
    - 2i\mu_{12}(\varepsilon_L^P)\rho_{12} \\
  & + i\mu_{23}(\varepsilon_L^C)^*\rho_{23}
    - i\mu_{23}(\varepsilon_L^C)\rho_{32}
    - \gamma_{31}(\rho_{33}-\rho_{11})
    - \gamma_{21}^D|\rho_{21}|^2, \\
  \dot{\rho}_{22} - \dot{\rho}_{33} =&  \notag
    - (2\gamma_{23} + \gamma_{21})\rho_{22}
    + i\mu_{12}(\varepsilon_L^P)^*\rho_{21}
    - i\mu_{12}(\varepsilon_L^P)\rho_{12} \\
  & + 2i\mu_{23}(\varepsilon_L^C)^*\rho_{23}
    - 2i\mu_{23}(\varepsilon_L^C)\rho_{32}
    - \gamma_{31}(\rho_{11}-\rho_{33})
    - \gamma_{23}^D|\rho_{23}|^2, \\
  \dot{\rho}_{23} =& \notag
    - i[\Delta_C - \epsilon_c(\rho_{22}-\rho_{33})]\rho_{23}
    - [\gamma - (\gamma_{23}^D/2)(\rho_{22} - \rho_{33})]\rho_{23} \\
  & + i\mu_{23}(\varepsilon_L^C)(\rho_{22}-\rho_{33})
    - i\mu_{12}(\varepsilon_L^P)\rho_{13}, \\
  \dot{\rho}_{21} =& \notag
    - i[\Delta_P - \epsilon_p(\rho_{22}-\rho_{11})]\rho_{21}
    - [\gamma - (\gamma_{21}^D/2)(\rho_{22} - \rho_{11})]\rho_{21} \\
  & + i\mu_{12}(\varepsilon_L^P)(\rho_{22}-\rho_{11})
    - i\mu_{23}(\varepsilon_L^C)\rho_{31}, \\
  \dot{\rho}_{31} =& \notag
    - [\gamma_{31} + i(\Delta_P - \Delta_C)]\rho_{31}
    -i[\epsilon_p (\rho_{22} - \rho_{11})
    -  \epsilon_c (\rho_{22} - \rho_{33})]\rho_{31} \\
  & - i\mu_{23}(\varepsilon_L^C)^*\rho_{21}
    + i\mu_{12}(\varepsilon_L^P)\rho_{32},
\end{align}
\end{subequations}\\[0.1cm]
where $\varepsilon_{L}^P$ and $\varepsilon _{L}^C$ are complex, microscopic, 
slowly-varying electric field envelopes of the probe and coupling fields, 
respectively; $\gamma_{21}$ and $\gamma_{23}$ are the radiative decay rates from 
level $\ket{2}$ to levels $\ket{1}$ and $\ket{3}$, respectively; population 
decays from level $\ket{3}$ back to level $\ket{1}$ at a rate $\gamma_{31}$ in 
non-radiative manner, and we define $\gamma = \half(\gamma_{21} + \gamma_{32} + 
\gamma_{31})$; $\Omega_P = 2\mu_{12}\,\varepsilon_{L}^P$ is the Rabi frequency 
of the probe field and ($\Omega_C = 2\mu_{32}\,\varepsilon_{L}^C$) the Rabi 
frequency of the coupling field; $\mu_{12}$ is the dipole matrix element for 
transition $\ket{1}\rightarrow\ket{2}$ and $\mu_{23}$ for transition 
($\ket{2}\rightarrow\ket{3}$). Equations~\eqref{Eq:DressedStates} describe the 
atomic dynamics of the system. $\gamma_{21}^D$ is the dipole-dipole induced 
decay rate due to NDD interaction to level $\ket{1}$ and $\gamma_{23}^D$ to 
level $\ket{3}$~\cite{Joshi19}.

The combined electric field acting on each atom inside the cavity is the sum of 
the probe and coupling fields, so that
\begin{equation}
 \Evec = (\Evec_P\,e^{-i\omega_P t} + \Evec_C\,e^{-i\omega_C t} + c.c.).
\end{equation}
The probe field $E_P$ at frequency $\omega_P$ (interacting with the atomic 
transition $\ket{1}\rightarrow\ket{2}$) circulates inside the cavity acting as 
the cavity field. The coupling field $E_C$ at frequency $\omega_C$ (coupling 
transition $\ket{3}\rightarrow\ket{2}$) does not circulate inside the optical 
ring cavity as shown in Fig.~\ref{Fig:ring}. However, the two fields propagate 
through the nonlinear medium in the same direction, so the first-order Doppler 
effect in this three-level, EIT system can be eliminated~\cite{Gea95:51}. In 
fact, the coupling field behaves like a controlling field. The field $E_P$ 
enters the cavity from the partially transparent mirror $M_1$ and drives the 
nonlinear medium. To determine the field inside the unidirectional ring cavity, 
we write the relations obtained by boundary conditions~\cite{Joshi10}
\begin{align}
  E_P^T &= \sqrt{T}\,E_P(L,t), \label{Eq:Polarization} \\
  E_P(0,t) &= \sqrt{T}\,E_P^I(t) + R\,e^{-i\delta_0}\,E_P(L,t-\Delta t),
  \label{Eq:Polarization1}
\end{align}
where the length of the atomic sample is specified by the parameter $L$, $\ell$ 
is the side arm length between $M_2$ and $M_3$ (and also $M_4$ and $M_1$), and 
the time taken by light to travel from mirror $M_2$ to $M_1$ via mirrors $M_3$ 
and $M_4$ is given by $\Delta t = (2\ell + L)/c$. The cavity frequency detuning 
is defined as $\delta_0 = (\omega_\text{cav} - \omega_P) L_T/c$, where 
$\omega_\text{cav}$ is the frequency of the nearest cavity mode to frequency 
$\omega_P$, and $L_T\approx2(\ell+L)$ represents the round-trip length of the 
ring cavity.

The dynamical evolution of the probe field inside the optical cavity is 
described by the Maxwell's equation (in the slowly-varying envelope 
approximation)~\cite{Joshi10}
\begin{equation}\label{Eq:Maxwell}
 \dfrac{\partial E_P}{\partial t} + c \dfrac{\partial E_P}{\partial z} =
 2\pi i\,\omega_P \mu_{12} P(\omega_P),
\end{equation} 
where $P(\omega_P) = N \mu_{21} \rho_{12}$ is the induced atomic polarization 
responsible for AOB, and $N$ is the number density of atoms. In order to obtain 
the polarization $P(\omega_P)$, one needs to first solve the set of 
density-matrix equations \eqref{Eq:DressedStates} in the steady-state limit, 
then integrate~\eqref{Eq:Maxwell} over the length of the atomic sample 
using~\eqref{Eq:Polarization} and~\eqref{Eq:Polarization1}. The steady-state 
boundary conditions can be written as~\cite{Joshi10}
\begin{align}\label{Eq:Polarization2}
  E_P^T &= \sqrt{T}\,E_P(L), \\
  E_P(0) &= \sqrt{T}\,E_P^I(t) + R\,e^{-i\delta_0}\,E_P(L).
\end{align}

\section{\label{sec:results}Results and Discussion}
\begin{figure}
  \includegraphics[scale=0.5]{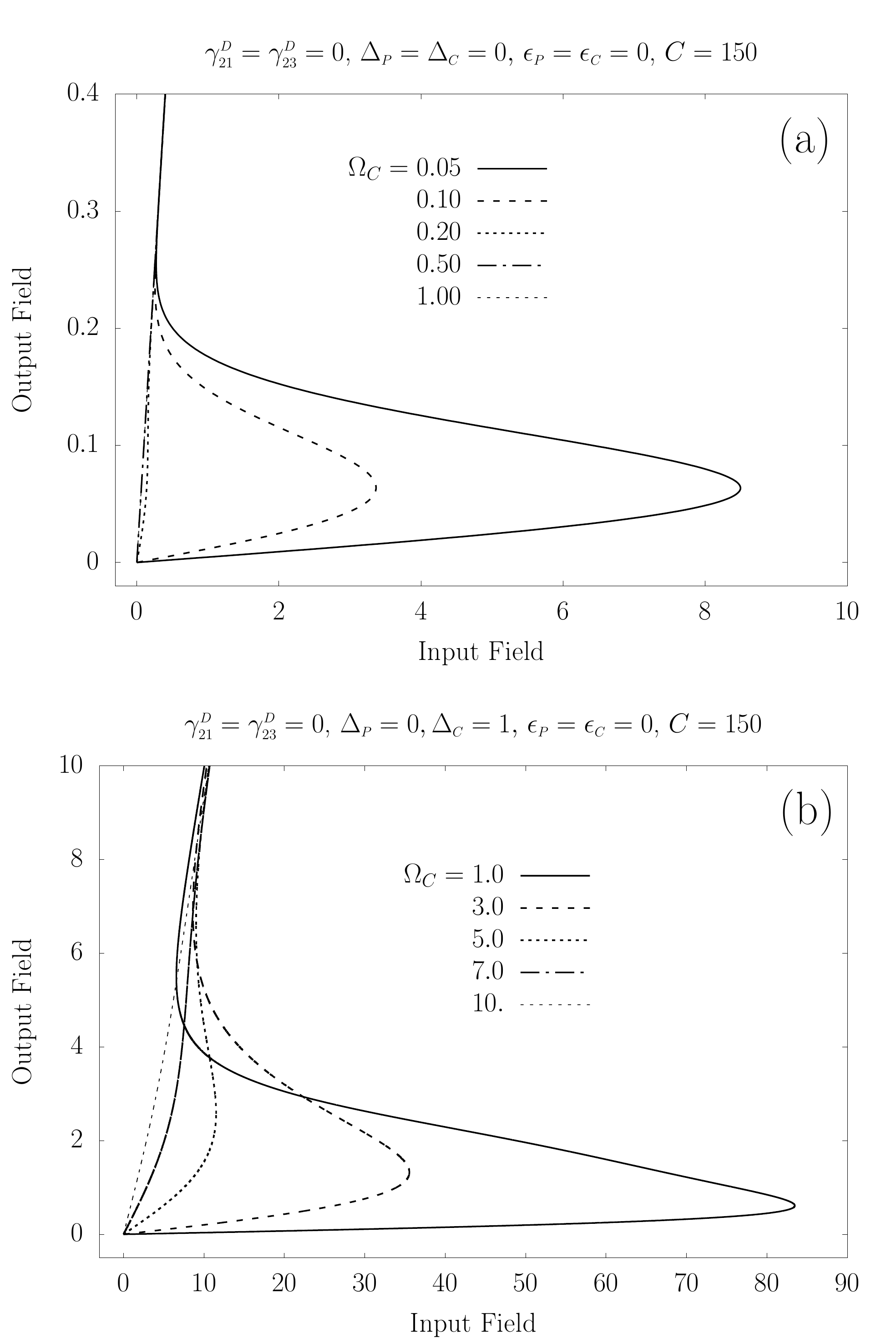}
  \caption{Optical bistability (in terms of output vs. input cavity field) for a
  three-level system in Raman configuration of its states. The Rabi frequency
  $\Omega_C$ is variable. (a) $\Delta_C=0$, (b) $\Delta_C = 1.0$.}
  \label{Fig:3}
\end{figure}

\begin{figure}
  \includegraphics[scale=0.5]{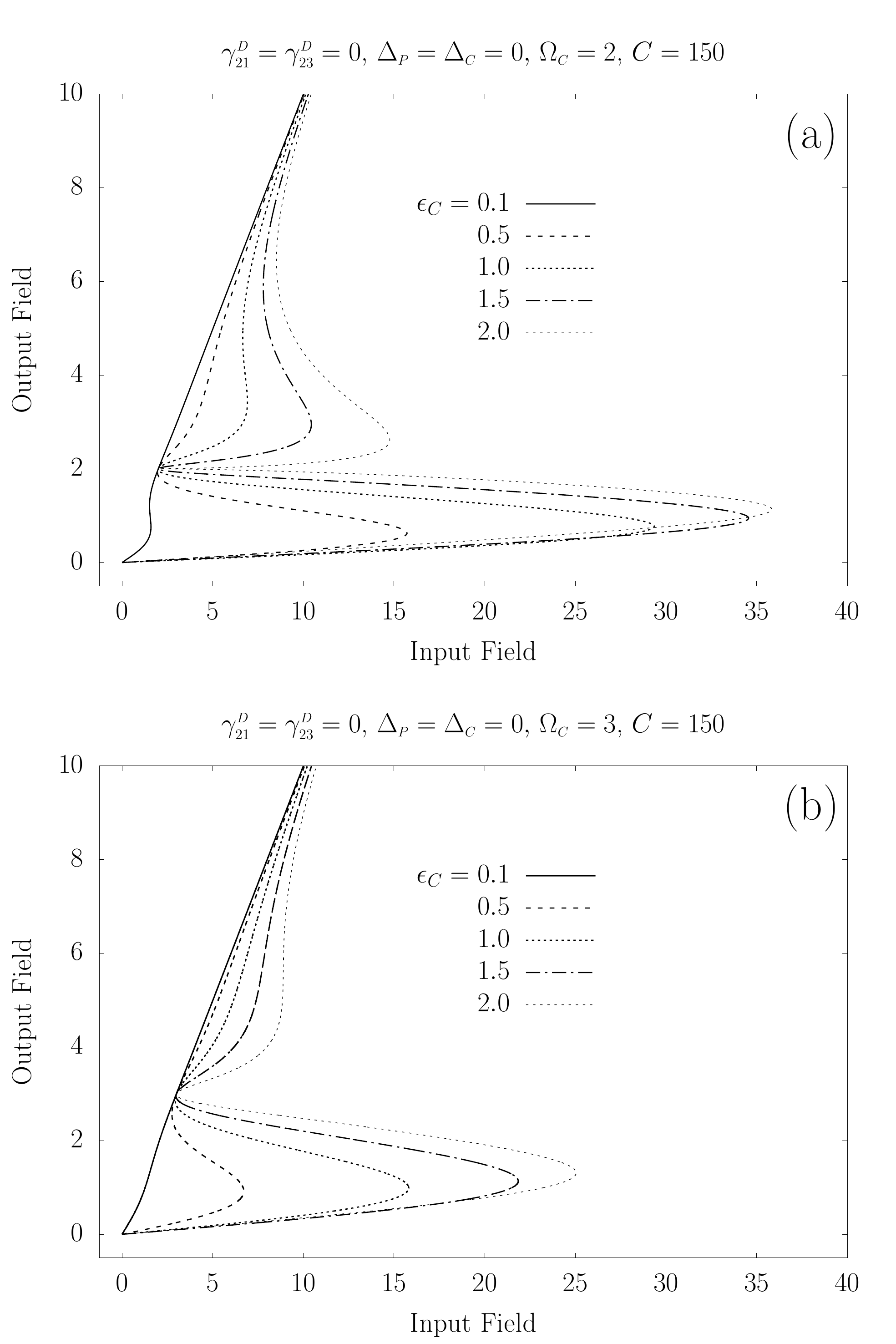}
  \caption{Output vs. input cavity field of the system as a result of non-zero,
  near dipole-dipole interaction (NDD). NDD (imaginary part) parameter
  $\epsilon_C$, which is varying. (a) $\Omega_C=2.0$, (b) $\Omega_C=3.0$.}
  \label{Fig:4}
\end{figure}

We solved equations~\eqref{Eq:DressedStates} numerically and plotted the cavity 
output field ($E_P^T$) versus the cavity input field ($E_P^I$) for various 
parametric values of the coupling field and NDD interaction parameter. The 
output-input features for a system showing OB are shown in 
Figs.~\ref{Fig:3}--\ref{Fig:6}.

Figure~\ref{Fig:3}(a) shows the curves for a three-level system in Raman 
configuration of its levels with no dipole-dipole interaction for different 
$\Omega_C$ values of the coupling field. The atomic sample is kept inside the 
cavity (see Fig.~\ref{Fig:ring}). The parameters used were 
$\gamma^{D}_{21}=\gamma^{D}_{23}=0$, $\Delta_P=\Delta_C=0$, 
$\epsilon_P=\epsilon_C=0$, and $C=150$ as the cooperative parameter for OB 
~\cite{Joshi10}. We changed $\Omega_C$ to $0.05$, $0.1$, $0.2$, $0.5$, and 
$1.0$. For a small $\Omega_C$, the OB threshold was large (solid line curve). 
When we increased $\Omega_C$, the threshold reduced, as seen in the figure. For 
a very large $\Omega_C$, the OB feature disappeared. The effect of a finite 
detuning of the coupling field is shown in Fig.~\ref{Fig:3}(b). Here, the only 
parameter modified was $\Delta_C=1.0$ leaving the other parameters unchanged. 
The non-zero coupling field detuning enhanced the OB threshold, as it can be 
observed by comparing the solid lines in Figs.~\ref{Fig:3}(a) 
and~\ref{Fig:3}(b). This threshold decreased for bigger values of the coupling 
field. The different curves correspond to values of $\Omega_C = 1.0$, $3.0$, 
$5.0$, $7.0$, and $10$ (see~\cite{Joshi10}).

Figures~\ref{Fig:4}(a,b) showed the impact of non-zero, real and imaginary parts 
of NDD interaction. We used parameters $\gamma^{D}_{21}= \gamma^{D}_{23}=0$, 
$\Delta_P=\Delta_C=0$, $\Omega_C=2.0$, and $C=150$. The effect of the imaginary 
part of NDD interaction is depicted in the curves plotted (Fig.~\ref{Fig:4}(a)) 
for $\epsilon_P=\epsilon_C=0.1$, $0.5$, $1.0$, $1.5$, and $2.0$. For smaller 
values of $\epsilon$ ($0.1$, $0.5$), the curves showed OB. For a further 
increase in the value of $\epsilon$ ($1.0$, $1.5$, $2.0$), we observed the 
interesting phenomenon of optical multistabilty in the curves. The threshold of 
all these OB and multistability curves can be altered by modifying the value of 
the coupling field because this changes the induced coherence and hence the 
threshold. One such interesting study is shown in Fig.~\ref{Fig:4}(b), where we 
changed $\Omega_C$ to $3.0$ but kept all other parameters the same. By doing so, 
the threshold of OB and multistability curves came down. Again, this was due to 
the altered induced coherence.

In Fig.~\ref{Fig:5}(a,b), we showed the controllability of OB by keeping the 
values of an NDD interaction parameter fixed. Parameters used for 
Fig.~\ref{Fig:5}(a) were $\gamma^{D}_{21}=\gamma^{D}_{23}=0$, 
$\Delta_P=\Delta_C=0$, $C=150$, and $\epsilon_C=\epsilon_P=1.0$. We plotted 
curves for $\Omega_C = 1.0$, $3.0$, $5.0$, $7.0$ and $10.0$. At lower values of 
$\Omega_C$ (1.0 and 3.0), we found multistability that gradually changed to OB 
as $\Omega_C$ further increased to 5.0, 7.0 and 10.0. Thus for a fixed value of 
the NDD parameter $\epsilon$ (imaginary part), the transition from OB to 
multistability can be controlled by changing the coupling field strength. 
However, when we introduced the real part of a NDD parameter, such as $\gamma 
^{D}_{21}= \gamma ^{D}_{23}=1.5$ in Fig.~\ref{Fig:5}(b), the shape of all curves 
remained similar to those in Fig.~\ref{Fig:5}(a) but the OB threshold of all 
curves increased. This effect is because of the associated increase in the 
system's dampening that reduces the induced coherence.

For Fig.~\ref{Fig:6}(a), the values for the NDD parameters were 
$\epsilon_C=\epsilon_P=2.0$, $\gamma ^{D}_{21} = \gamma ^{D}_{23}=0.0$, while 
for Fig.~\ref{Fig:6}(b), the values were $\epsilon_C=\epsilon_P=2.0$, and 
$\gamma ^{D}_{21}= \gamma^{D}_{23}=3.0 $. Figures~\ref{Fig:6}(a,b) showed 
reduction in the multistability structure for $\Omega_C=1.0$ when compared with 
Fig.~\ref{Fig:5}(a,b). $\Omega_C=3.0$ still showed multistability, but the 
threshold raised as compared to those in Figs.~\ref{Fig:5}(a,b). For further 
increase in $\Omega_C$, all curves displayed OB but with an augmented threshold 
value. The physical interpretation of these results is implicit in the 
manipulation of the physical properties of the three-level, EIT medium 
considered in this work. We achieved this by varying the coupling field and the 
NDD interaction parameters in the atomic system that changed its generated 
atomic coherence~\cite{Joshi19}. While keeping the coupling field constant, we 
varied the NDD parameters to control absorption, dispersion, and nonlinearity of 
the medium and to get changes in its OB.

\section{\label{sec:conclusion}Conclusion}

In this work, we studied the optical bistability displayed by a three-level, 
$\Lambda$-type electromagnetic induced transparency medium including near 
dipole-dipole interaction. The inclusion of NDD interaction provided another 
control parameter for changing the absorption, dispersion, and nonlinearity of 
such EIT system. In fact, by modifying the number density of the atomic medium, 
we may obtain wanted values of NDD parameters. Consequently, by the combined 
effect of the coupling field and NDD interaction parameters, the system exhibits 
interesting changes as it passes from OB to optical multistability. This results 
could be experimentally realized in the three-level atomic system described in 
Refs.~\cite{Joshi06:49,Joshi10}.

\section*{Acknowledgments}
One of us (A.J.) is thankful to G. Duree (RHIT) for his interest and support 
for this work.


\newpage
\section*{Figure Captions}

\noindent
{\bf Figure 1:} Schematic diagram of a unidirectional ring cavity having four
  mirrors ($M_1\text{--}M_4$) and an atomic vapor cell of length $L$. Mirrors
  $M_3$ and $M_4$ are perfectly reflecting mirrors ($R=1$). The incident and
  transmitted fields are represented by $E_P^I$ and $E_P^T$, respectively,
  while the coupling field is represented by $E_C$. \\

\noindent
{\bf Figure 2:} Schematic diagram of the three-level system in a
  $\Lambda$-configuration driven by probe and coupling lasers of frequencies
  $\omega_P$ and $\omega_C$, respectively. \\
  
\noindent
{\bf Figure 3:} Optical bistability (in terms of output vs. input cavity field) 
  for a three-level system in Raman configuration of its states. The Rabi 
  frequency $\Omega_C$ is variable. (a) $\Delta_C=0$, (b) $\Delta_C = 1.0$. \\

\noindent
{\bf Figure 4:} Output vs. input cavity field of the system as a result of
  non-zero, near dipole-dipole interaction (NDD). NDD (imaginary part) parameter
  $\epsilon_C$, which is varying. (a) $\Omega_C=2.0$, (b) $\Omega_C=3.0$. \\

\noindent
{\bf Figure 5:} Output vs. input cavity field of the system for non-zero, near
  dipole-dipole interaction with $\epsilon_P=\epsilon_C=2.0$. The Rabi frequency
  $\Omega_C$ is variable. NDD (real part) parameters $\gamma_{21}^D$ and
  $\gamma_{23}^D$ are changed as: (a) $\gamma_{21}^D=\gamma_{23}^D=0$,
  (b) $\gamma_{21}^D=\gamma_{23}^D=1.5$. \\

\begin{figure}
  \includegraphics[scale=0.5]{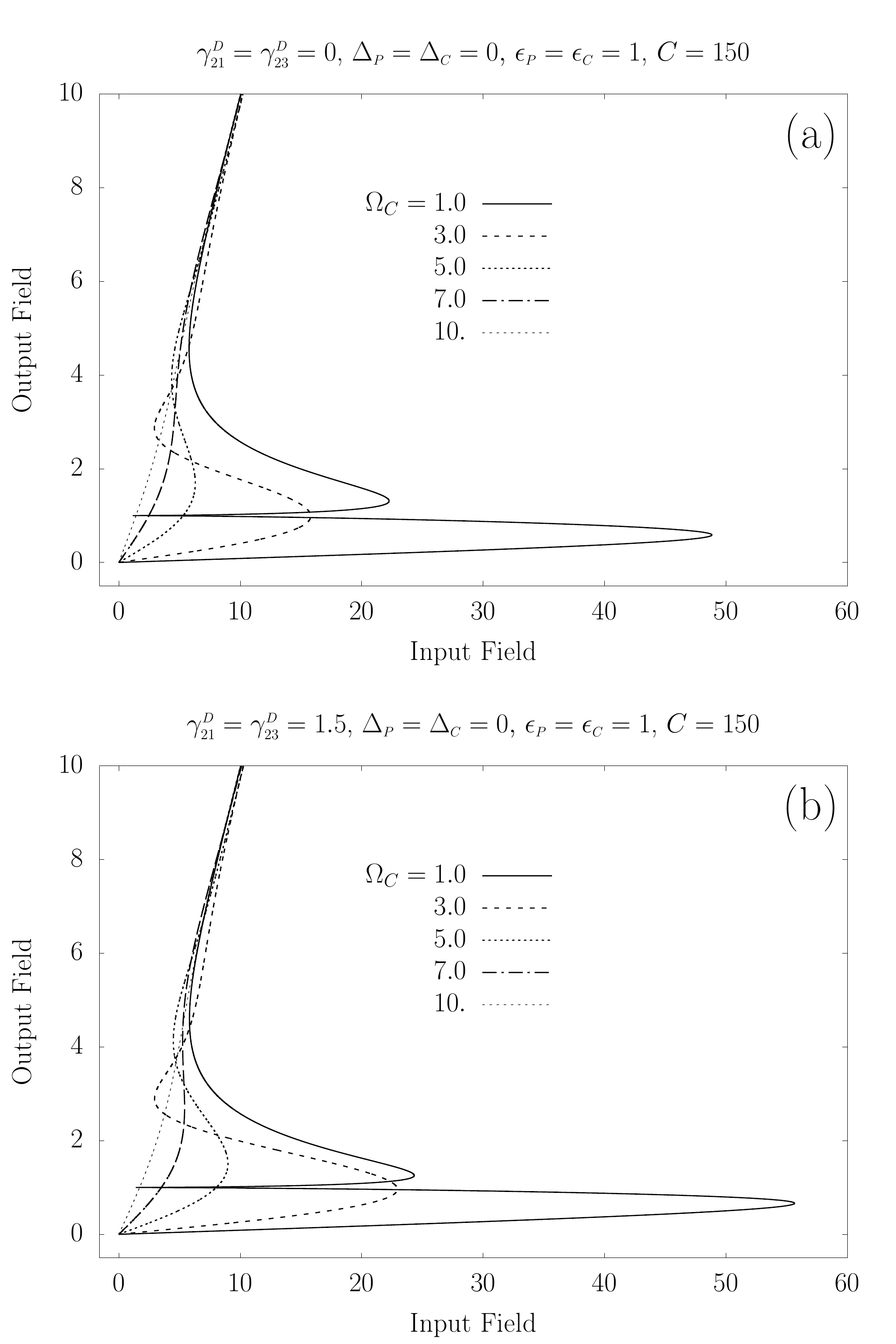}
  \caption{Output vs. input cavity field of the system for non-zero, near
  dipole-dipole interaction with $\epsilon_P=\epsilon_C=2.0$. The Rabi frequency
  $\Omega_C$ is variable. NDD (real part) parameters $\gamma_{21}^D$ and
  $\gamma_{23}^D$ are changed as: (a) $\gamma_{21}^D=\gamma_{23}^D=0$,
  (b) $\gamma_{21}^D=\gamma_{23}^D=1.5$.}
  \label{Fig:5}
\end{figure}

\begin{figure}
  \includegraphics[scale=0.6]{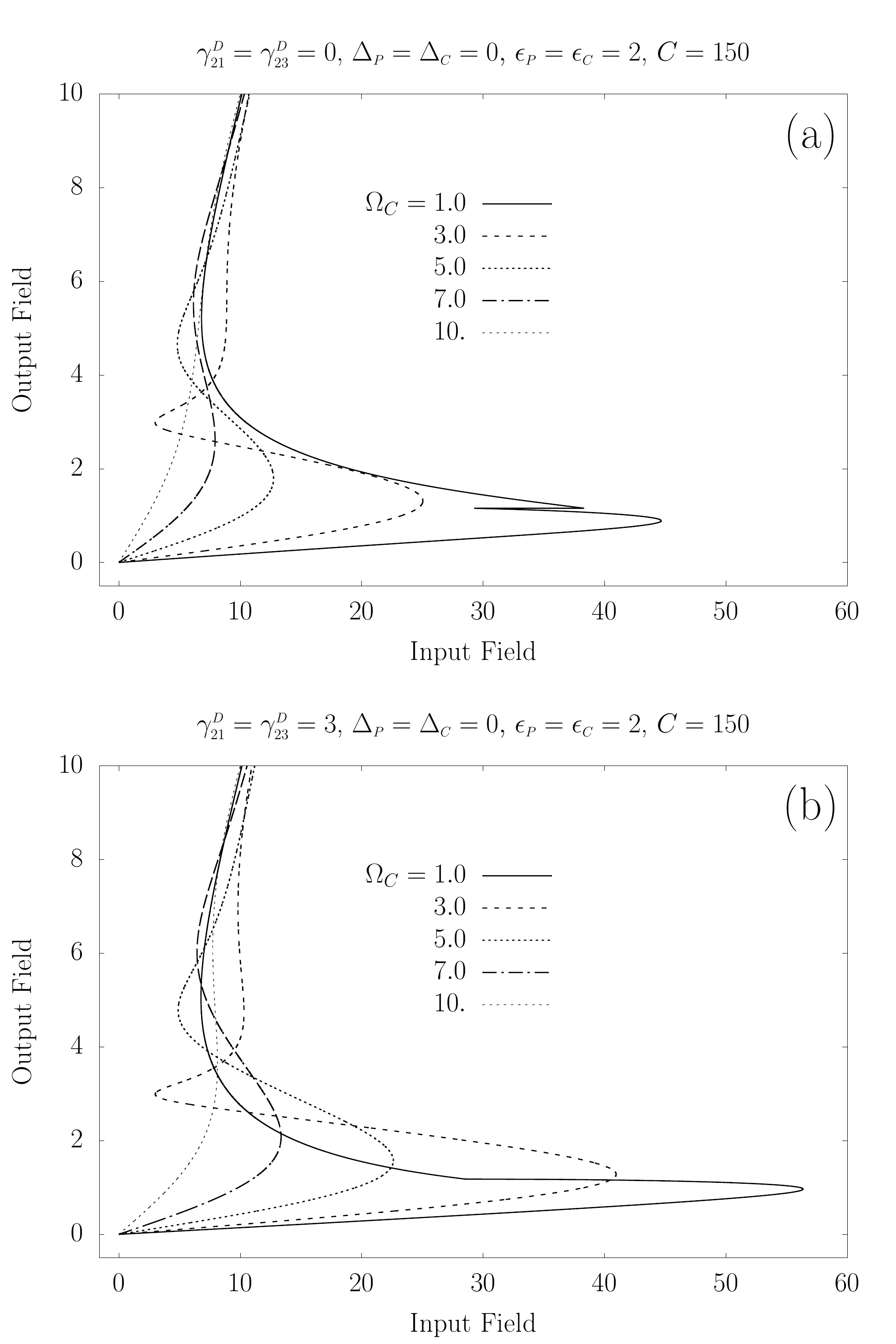}
  \caption{Same as Fig.~\ref{Fig:5}, with $\epsilon_P=\epsilon_C=2.0$.
  (a) $\gamma_{21}^D=\gamma_{23}^D=0$, (b) $\gamma_{21}^D=\gamma_{23}^D=3.0$.}
  \label{Fig:6}
\end{figure}

\noindent
{\bf Figure 6:} Same as Fig.~\ref{Fig:5}, with $\epsilon_P=\epsilon_C=2.0$.
  (a) $\gamma_{21}^D=\gamma_{23}^D=0$, (b) $\gamma_{21}^D=\gamma_{23}^D=3.0$. \\

\newpage
 
\clearpage

\clearpage

\clearpage

\clearpage

\clearpage
%
%
%

\end{document}